\documentclass[
twocolumn,
superscriptaddress,
 amsmath,amssymb,
 aps,
nofootinbib]{revtex4-1}

\usepackage{graphicx}
\usepackage{dcolumn}
\usepackage{bm}
\usepackage[dvipsnames]{xcolor}
\usepackage{soul}
\usepackage{ulem}
\usepackage{xr}
\usepackage{braket}
\usepackage{multirow}
\definecolor{green}{HTML}{008000}
\definecolor{darkpurple}{HTML}{7B39A7}

\usepackage[T1]{fontenc}
\usepackage{lmodern} 
\usepackage[utf8]{inputenc}

\begin{document}

\title{Unraveling real-time chemical shifts in the ultrafast regime}

\author{Daniel E. Rivas,\textsuperscript{1,*,$\dagger$} Lorenzo Paoloni,\textsuperscript{2,3,4,*} Rebecca Boll,\textsuperscript{1} Alberto De Fanis,\textsuperscript{1} Ana Mart\'inez Guti\'errez,\textsuperscript{2,3} Tommaso Mazza,\textsuperscript{1} Sol\`{e}ne Oberli,\textsuperscript{2,5,6} Oliver Alexander,\textsuperscript{7} Andr\'e Al-Haddad,\textsuperscript{8} Thomas M. Baumann,\textsuperscript{1} Christoph Bostedt,\textsuperscript{5,8} Simon Dold,\textsuperscript{1} Gianluca Geloni,\textsuperscript{1} Markus Ilchen,\textsuperscript{1,9,10,11} Dooshaye Moonshiram,\textsuperscript{4} Daniel Rolles,\textsuperscript{12} Artem Rudenko,\textsuperscript{12} Philipp Schmidt,\textsuperscript{1} Svitozar Serkez,\textsuperscript{1} Sergey Usenko,\textsuperscript{1} \'Angel Mart\'in Pend\'as,\textsuperscript{13} Michael Meyer,\textsuperscript{1} Jes\'us Gonz\'alez-V\'azquez,\textsuperscript{2,$\ddagger$} and Antonio Pic\'on\textsuperscript{2,3,4,$\mathsection$}\\
\  \\
\small{\textit{\textsuperscript{1}European XFEL, Holzkoppel 4, 22869 Schenefeld, Germany}\\
\textit{\textsuperscript{2}Departamento de Qu\'{i}mica, Universidad Aut\'{o}noma de Madrid, 28049 Madrid, Spain}\\
\textit{\textsuperscript{3}Condensed Matter Physics Center (IFIMAC), Universidad Autónoma de Madrid, 28049, Madrid, Spain}\\
\textit{\textsuperscript{4}Instituto de Ciencia de Materiales (ICMM-CSIC), 28049, Madrid, Spain}\\
\textit{\textsuperscript{5}Laboratory for Ultrafast X-ray Sciences, Institute of Chemical Sciences and Engineering, \'Ecole Polytechnique F\'ed\'erale de Lausanne (EPFL), CH-1015 Lausanne, Switzerland}\\
\textit{\textsuperscript{6}Laboratory of Theoretical Physical Chemistry, Institute of Chemical Sciences and Engineering, \'Ecole Polytechnique F\'ed\'erale de Lausanne (EPFL), CH-1015 Lausanne, Switzerland}\\
\textit{\textsuperscript{7}Department of Physics, Blackett Laboratory, Imperial College London, SW7 2AZ London, U.K.}\\
\textit{\textsuperscript{8}Paul-Scherrer Institute, CH-5232, Villigen PSI, Switzerland}\\
\textit{\textsuperscript{9}Institut für Experimentalphysik, Universität Hamburg, Luruper Chaussee 149, 22761 Hamburg, Germany}\\
\textit{\textsuperscript{10}Deutsches Elektronen-Synchrotron DESY, Notkestr. 85, 22607 Hamburg, Germany}\\
\textit{\textsuperscript{11}Center for Free-Electron Laser Science CFEL, Deutsches Elektronen-Synchrotron DESY, Notkestr. 85, 22607 Hamburg, Germany}\\
\textit{\textsuperscript{12}J.R. Macdonald Laboratory, Department of Physics, Kansas State University, Manhattan, Kansas 66506, USA}\\
\textit{\textsuperscript{13}Department of Analytical and Physical Chemistry, University of Oviedo, E-33006, Oviedo, Spain}}}

\footnotetext{These authors contributed equally}
\footnotetext{Corresponding author, daniel.rivas@xfel.eu}
\footnotetext{Corresponding author, jesus.gonzalezv@uam.es}
\footnotetext{Corresponding author, antonio.picon@csic.es}

\date{\today}
\begin{abstract}

Traditional x-ray photoelectron spectroscopy (XPS) relies upon a direct mapping between the photoelectron binding energies and the local chemical environment, which is well-characterized by an electrostatic partial charges model for systems in equilibrium. However, the extension of this technique to out-of-equilibrium systems has been hampered by the lack of x-ray sources capable of accessing multiple atomic sites with high spectral and temporal resolution, as well as the lack of simple theoretical procedures to interpret the observed signals. In this work we employ XPS with a narrowband femtosecond x-ray probe to unravel different ultrafast dissociation processes of a polyatomic molecule, fluoromethane (CH$_{3}$F). We demonstrate that the PC model can be successfully applied to describe the C-F and C-H dissociation dynamics after strong-field ionization, with excellent agreement between experimental measurements and ab initio simulations.These results enable the application of this technique to out-of-equilibrium systems of higher complexity, by correlating real-time information from multiple atomic sites and interpreting the measurements through a viable theoretical modeling.

\end{abstract}

\maketitle


X-ray radiation can remove electrons from localized atomic-core orbitals through photoionization. The binding energy of these electrons exhibit differences depending on the local chemical environment, as realized in early Nobel-prize-winning studies of x-ray photoelectron spectroscopy (XPS) with high energy resolution \cite{siegbahn_electron_1982}. These so-called chemical shifts depend on the surrounding electron density distribution and on the nuclear geometry \cite{gelius_binding_1974}. Nowadays, XPS is a well-established technique in a broad range of applications for systems in the (quasi-) static domain, especially in Materials Sciences \cite{siegbahn_esca_1969,lovelock_photoelectron_2010,greczynski_x-ray_2020,bagus_x-ray_2024}. 

Chemical shifts are well described by partial charges (PC) models in the form \cite{gelius_binding_1974}:
\begin{eqnarray}
\Delta E_a = k q_a + \sum_{b\neq a} \frac{q_b}{R_{ab}} + l
\label{eq:partialchargemodel}
\end{eqnarray}
where $\Delta E_a$ is the binding energy shift of the atom being ionized $a$,  $q_a$ is the partial charge of that atom, $k$ is a constant that depends on the Coulomb interaction between the core orbital and the valence orbitals \cite{davis_x-ray_1972}, $q_b$ are the partial charges at the other neighboring atoms $b$, $R_{ab}$ are the distances between the ionized atom and the other atoms, and $l$ is the energy reference. Through this model we understand that a chemical shift will depend on both the local charge at the atom being ionized, as well as the charge from the neighboring atoms. This model is derived for the ground state in the mean-field approximation. Therefore, its validity in out-of-equilibrium systems, e.g. during the formation or cleavage of chemical bonds, represents a fundamental question for applications in ultrafast science. This question gained new relevance due to the advent of x-ray free-electron laser (XFEL) facilities, which open the possibility to produce few-femtosecond pulses in the x-ray regime. With these tools, the XPS technique can be extended to investigate dynamics of out-of-equilibrium systems, in particular photo-induced reactions \cite{lennox_excited-state_2017,chen_amino_2018,pathak_tracking_2020,kang_intrinsic_2002,cheng_dynamics_2009,polli_conical_2010,chergui_ultrafast_2015}.

\begin{figure*}\centering\includegraphics[scale=0.4]{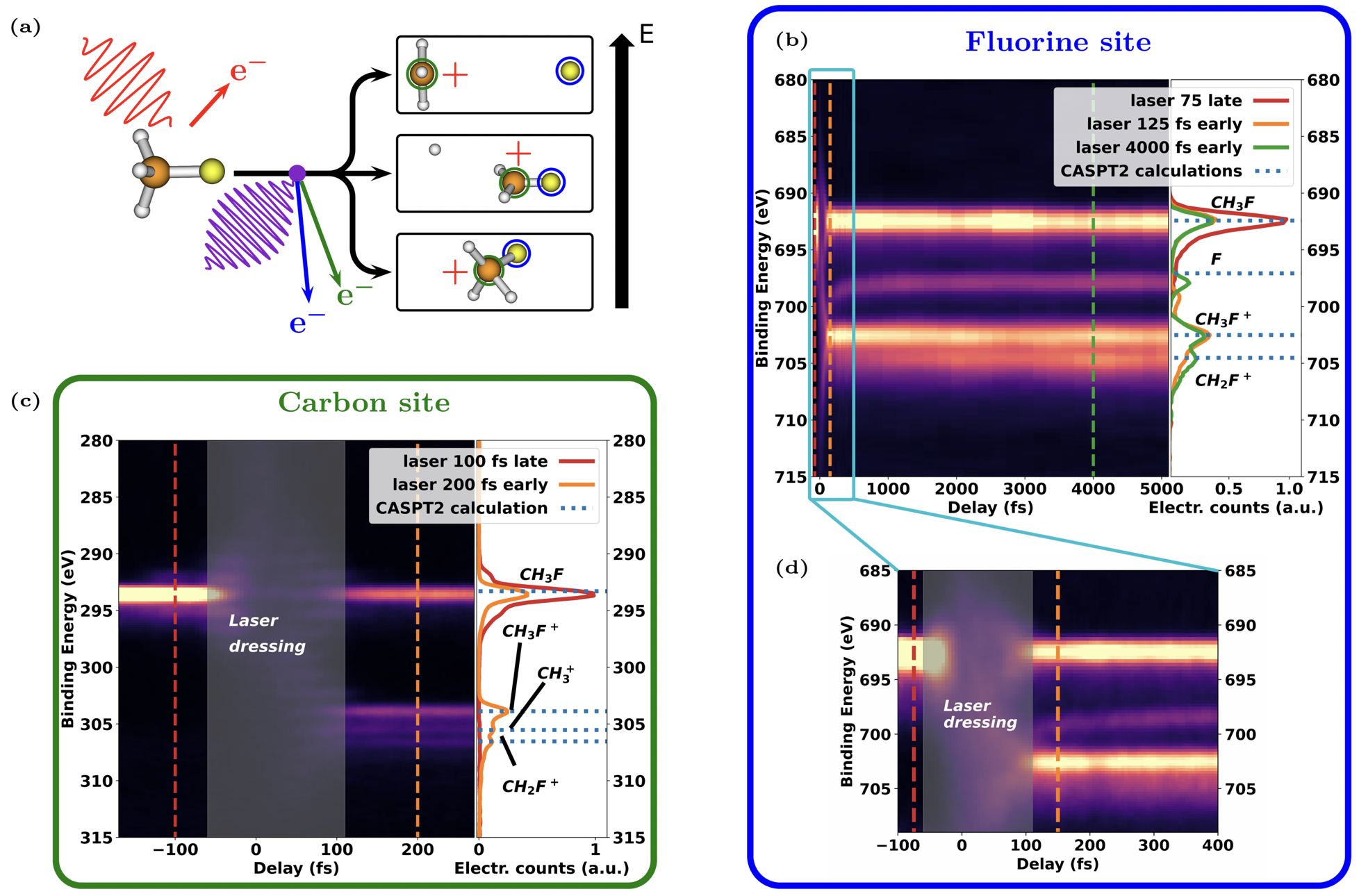} 
\caption{{\bf Time-resolved XPS to investigate the ultrafast ionization dynamics of CH$_{3}$F  }. a) Illustration of the main transient state channels in CH$_{3}$F after IR strong-field ionization, corresponding to a stable CH$_{3}$F$^+$ cation (bottom panel), a CH$_2$F$^+$ and H dissociation (middle panel), or a CH$_3^+$ and F dissociation (top panel). After the IR ionization step, a femtosecond x-ray pulse ionizes the 1s electrons and enables to investigate the local chemical shifts at the F and C sites (blue and green arrows and circles respectively). The vertical energy bar illustrates the relative energy of the channels in increasing order. b) and c) Experimental time-resolved XPS traces at the F and C sites, respectively, and corresponding lineouts at selected time delays. Positive delays corresponds to the laser pump pulse arriving earlier than the x-ray probe pulse. The laser-dressing region between -60 and 110 fs is highlighted in grey in (c). The measured peaks are identified by using a high-precision level of theory (dashed blue lines, see Methods) and are shown in table \ref{table:BEs}. A shift of 0.3 eV has been applied to the calculated lines in the lineouts panel (c) as explained in section S2 of the SM. (d) Time-resolved XPS traces for the F site, zoomed into the -100 to 400 fs delay region. The laser dressing region between -60 fs and 110 fs is highlighted in grey. }
\label{fig:scheme}
\end{figure*}

Correlating chemical shifts with specific dynamics in an evolving system remains a significant challenge, as both electronic and nuclear contributions can influence the observed shifts. In general, ab-initio calculations of core-level binding energies are essential for interpreting experimental data. While current computational methods are capable of simulating nuclear dynamics and tracking charge redistribution in excited-state systems, the accurate calculation of inner-shell binding energies is considerably more demanding. This is primarily due to the need to explicitly model core-hole states, which are computationally intensive and require specialized approaches beyond standard electronic structure methods \cite{neville_ultrafast_2018,inhester_spectroscopic_2019}.

Recent theoretical and experimental works have confirmed that strong chemical shifts occur after core-hole \cite{al-haddad_observation_2022} or valence-band excitations \cite{leitner_time-resolved_2018, brause_time-resolved_2018,mayer_following_2022,allum_localized_2022,gabalski_time-resolved_2023} , either due to the excited states involved or the cleavage of a bond.  The majority of photo-induced processes involve multiple intertwined electronic and nuclear dynamics, making it challenging to directly link the observed spectral shifts to specific underlying mechanisms. Recent advances in spectral resolution have enabled clearer separation of these processes, allowing distinct channels to be resolved and analyzed individually \cite{Facciala2025, Thomson2025}.

When extending high-resolution XPS to increasingly complex dynamics, the interpretation would largely benefit from employing a PC model, such as the one given by Eq. (\ref{eq:partialchargemodel}). This approach requires only the simulation of partial charge evolution, significantly reducing computational complexity. A first step in this direction was given in Ref. \cite{mayer_following_2022}, where they extended the traditional interpretation of chemical shift to include transient excited states, in which an electronic transition involving a nonbonding orbital has strong local charge effect.

Here we investigate dynamics in fluoromethane after strong-field ionization with time-resolved XPS (tr-XPS) and show that the obtained high-resolution experimental results are described through a simple extension of the PC model to the ultrafast regime. An intense 800-nm pulse ionizes the molecule which either breaks at the C-F bond on femtosecond time-scales or stays in a partially stable state, possibly  exhibiting a C-H bond cleavage within a picosecond time-window. To capture these dynamics, we measure the real-time chemical shift at the K-shell of both the fluorine and carbon atomic sites (approximately 690 eV and 280 eV binding energy, respectively). We show that due to the change in ionic state, the binding energies of the cations are well-separated from those of the neutral molecule, which allows their tracking in time. 
Our findings validate the use of the PC model for interpreting tr-XPS experimental data. We show that the partial charges obtained from a Mulliken population analysis account for the observed chemical shifts. The results from the PC model manifest the significant role of charges located on atoms neighboring the x-ray absorbing site. Notably, this effect can remain significant even when the contributing atoms are located at considerable distances from the absorbing atom.

\section*{Ultrafast chemical shifts}

Our experiment was performed at the Small Quantum Systems (SQS) scientific instrument of the European X-ray free-electron laser (EuXFEL) \cite{mazza_beam_2023}. Femtosecond optical-laser pump pulses are overlapped with monochromatized x-ray pulses of 928.3 eV photon energy and 0.33 eV bandwidth (FWHM) resulting in an overall temporal resolution of 35 fs FWHM. The interaction occurs at the Atomic-like Quantum Systems (AQS) end-station equipped with an electron time-of-flight spectrometer (see Methods for additional information). At the interaction point, fluoromethane molecules delivered through effusive expansion from a needle are strong-field ionized by the optical laser, which predominantly populates the ground state and first two excited states of the cation (see section S3-C  of the SM). The ground and first excited state are degenerate in the Frank-Condon region, and exhibit a slight barrier, either keeping the cation bound or permitting C-H bond cleavage [see Fig. \ref{fig:scheme}(a)]. The second excited state is dissociative along the C-F bond. The dynamics are captured through the x-ray probe pulse, producing photoelectrons which are analyzed via an electron time-of-flight spectrometer \cite{de_fanis_high-resolution_2022}. The x-ray photon energy is well above the ionization K-shell of carbon ($\sim$280~eV) and fluorine ($\sim$690~eV) and the spectrometer retardation is tuned accordingly to achieve high resolution for either carbon or fluorine photoelectrons.

The photoelectron binding energies (BE) as a function of the time delay between the two pulses, i.e. the tr-XPS trace, are shown in Figs. \ref{fig:scheme}(b) and (c) for the carbon and fluorine K-shells. For negative time delays (x-ray probe pulses before the optical pump pulse), a single main peak is measured, which corresponds to the K-shell ionization of the CH$_3$F ground state (293.56 eV and 692.4 eV from the C and F sites, respectively \cite{fukuzawa_site-selective_2007,thomas_x-ray_1970}).  For positive time delays, there is a clear depletion of the ground-state signal and new features appear at higher binding energy, which are related to the transient-state cation dynamics. From the depletion we determine that 65$\%$ of the molecules were ionized in the probed region. Because of the change in ionic state, a large increase in the binding energy of $\sim10$~eV compared to the neutral species is observed in both sites. We identify these peaks by performing high-accuracy static ab-initio calculations \cite{andersson_secondorder_1992,prigogine_new_1996} of the XPS spectrum for the products of the different channels, as shown in the line-out panels of Figs. \ref{fig:scheme}(b) and (c) and summarized in table \ref{table:BEs} (see Methods and section S2 in the SM for more information about the theoretical calculations). This enables us to discern two main dissociative channels, i) CH$_3$F$^+ \rightarrow$ CH$_3^+$+F and ii) CH$_3$F$^+ \rightarrow$ CH$_2$F$^+$+H, which are discussed in the next section. Finally, in the region close to the pump and probe temporal overlap, a strong smearing of the photolines is observed due to laser dressing (highlighted in gray in Figs. \ref{fig:scheme}(b) and (d), respectively). Due to the high intensities involved, this effect occurs despite the laser having a close to perpendicular polarization to the x-rays (see Methods).

\begin{table}
\begin{center}
\begin{tabular}{|c| c| c| } 
 \hline 
 Fragment & C K-shell BE (eV) & F K-shell BE (eV) \\ [0.5ex] 
  \hline
 CH$_3$F & 293.1 & 692.1  \\ 
 \hline

CH$_3$F$^+$ &303.6 & 702.1  \\ 
 \hline
 CH$_2$F$^+$ & 306.2 & 704.2  \\
 \hline
 CH$_3^+$ & 305.2 & -  \\
 \hline
 F & - & 696.8  \\ 
 \hline
\end{tabular}
\end{center}
\caption{ Calculated binding energies of the ground state of CH$_3$F and the different fragments produced after strong-field ionization.}
\label{table:BEs}
\end{table}

\begin{figure*}\centering\includegraphics[scale=1.3]{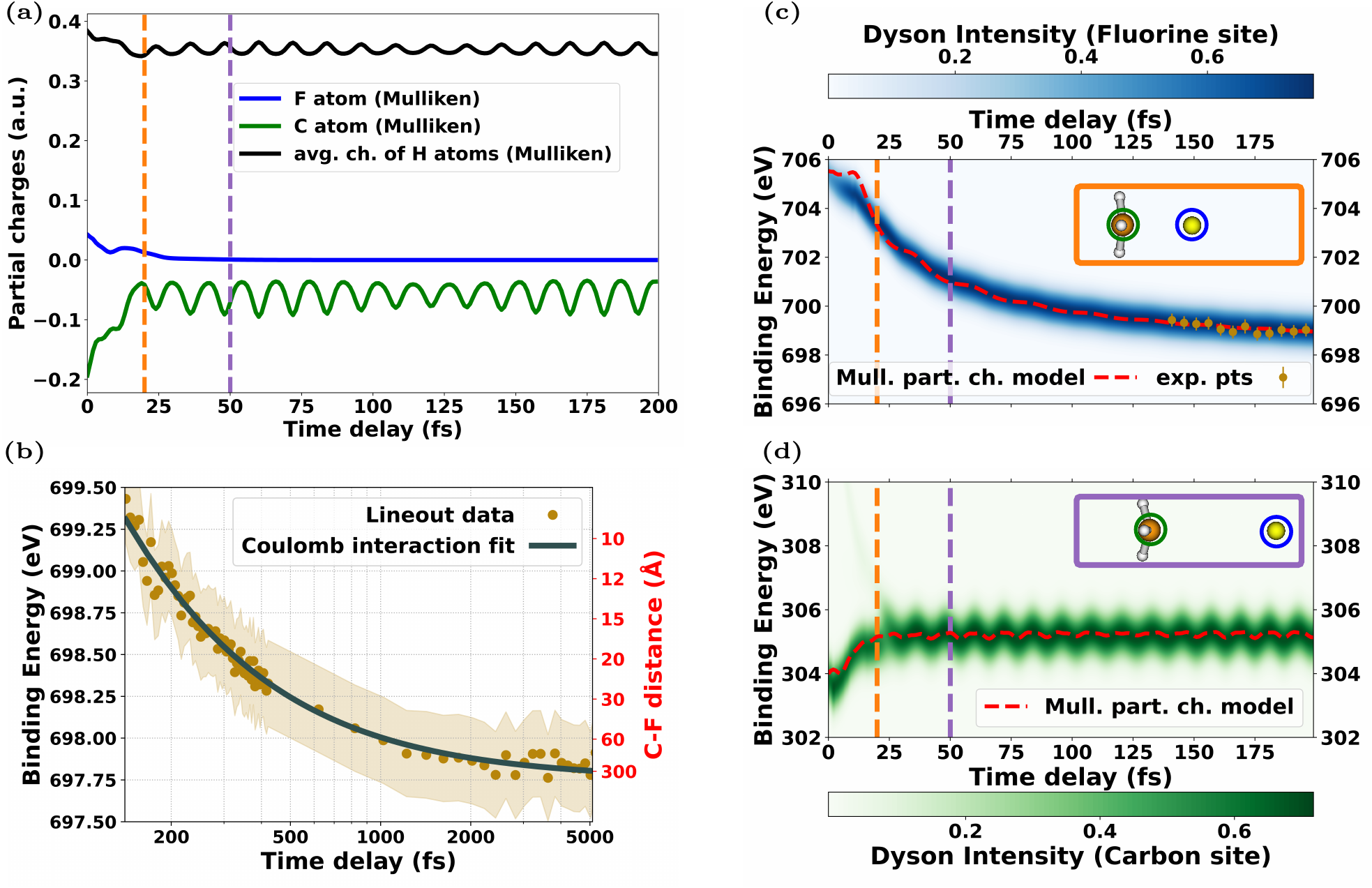}
\caption{{\bf Dynamical chemical shift during C-F bond elongation}. a) Calculated partial charges (Mulliken charges) at the location of the F, C and H atoms, for a single C-F dissociation trajectory. b) Measured time-dependent chemical shift of atomic F (golden dots, shaded region corresponds to the standard deviation) and fitted function based on the Coulomb interaction model (grey line), described in the main text. (c) and (d) Calculated binding energy based on PC model applied with the calculated Mulliken charge (red line) and Dyson intensity (blue colorscale for F and green for C), calculated through the ab-initio model including core-hole states. A shift of 0.5 eV is applied to the calculations in panel (c) as explained in section S3-C of the SM.  Dotted lines in (a), (c), and (d) mark the delays for the representative snapshots shown in the inset of (c) and (d). A weak satellite feature between 306 and 310 eV is observed in (d) for delays below 25 fs. }
\label{fig:CF}
\end{figure*}

\section*{Fluorine loss and long-range effects on chemical shifts}

A distinctive time-varying chemical shift is immediately apparent between $\sim$697-700 eV in the tr-XPS trace of the F K-shell [Figs. \ref{fig:scheme}(b) and (d)], which does not have a corresponding dynamical feature at the C K-shell [Fig. \ref{fig:scheme}(c)]. We attribute this feature to the ultrafast CH$_3$F$^+ \rightarrow$ CH$_3^+$+F dissociation, based on the binding energy calculations presented in table \ref{table:BEs}. To corroborate this, and check the applicability of Eq. (\ref{eq:partialchargemodel}) for this process, we perform molecular dynamics simulations aiming at determining the electron density distribution during the dissociation process (see Methods). We use a semi-classical surface-hopping approach, in which the electronic structure is treated quantum mechanically while the nuclear propagation is treated classically. For this particular channel, the second excited state of the cation is considered, with a BE at 17 eV from the ground state, presenting a dissociative behavior along the C-F bond (see the calculated potential energy curves in Fig. S9 of the Supplementary Material).

Fig. \ref{fig:CF}(a) shows the calculated Mulliken charges as the partial charge evolution for a selected trajectory. We observe that within a few tens of femtoseconds the partial charge on the F site stabilizes and remains constant thereon, hence we expect that the only remaining time-dependent term in Eq. (\ref{eq:partialchargemodel}) is the Coulomb interaction with the detached CH$_3^+$ cation. Therefore, the dynamic binding energy can be modeled as $ \mathrm{E_b}(t) = \mathrm{E_{b,0}} + 1/R_{\mathrm{CF}}(t)$, where $\mathrm{E_b}(t)$ is the instantaneous binding energy at a given distance to the CH$_3^+$ cation, $\mathrm{E_{b,0}}$ is the binding energy of the isolated neutral fluorine atom and $R_{\mathrm{CF}}(t)$ is the interatomic distance between the F atom and the CH$_3^+$ cation. We perform a fitting routine to extract the time-dependent atomic binding energy of F from the measured tr-XPS trace and fit it with the simple Coulomb model, as shown in figure \ref{fig:CF}(b). We observe a good agreement between the experimental data and the fitted Coulomb model for all recorded delays between 125 fs and 5 ps - the range at which the partial charge on the F is already constant (see Fig. \ref{fig:CF}(a)). Shorter delays are obscured by the laser dressing effect.  

From the fitted model we can also extract the real-time interatomic distance between the fluorine atom and the methyl ion, displayed on the right axis of figure \ref{fig:CF}(b). It is worth noting the high sensitivity of the F binding energy to charges at distances of tens of Ångströms, suggesting the future application of this technique to resolve dynamics in more complex macromolecules or nanoscopic systems. Finally, from this fitting procedure we determine a binding energy of $\mathrm{E_{b,0}} = 697.8 \pm 0.1\, \mathrm{eV}$ for isolated fluorine atoms, which, to the best of our knowledge, has not been reported in literature so far. This value differs from the one calculated in table \ref{table:BEs}, which is expected when not considering a full energy relaxation in the calculation (see section S2 in the SM).

We now turn to compare the results of the PC model in Eq. (\ref{eq:partialchargemodel}) with ab-initio calculations of the real-time binding energies. For the latter, we calculate the instantaneous binding energies at each time step of the molecular dynamic calculation, by taking the energy difference between the corresponding electronic state and the core-hole state (see Methods). The instantaneous binding energies calculated by applying the PC model using the Mulliken charges [red lines in Fig. \ref{fig:CF}(c) and (d)] show a good agreement with the ab-initio simulations [blue and green shaded plots for F and C in Fig. \ref{fig:CF}(c) and (d), respectively], for delays beyond 25 fs. At shorter time delays there is a slight deviation between the models, but as we show in the next section, it is possible to find a perfect agreement between them by slightly modifying the Mulliken charges. 

On the F site, the long-range pure Coulomb interaction is confirmed for longer time-delays. At delays shorter than 50 fs the interaction with the partial charges of the other atomic species become relevant, reflected in the dampened oscillation with a 24 fs period. This is due to the  umbrella motion of the CH$_3$ moiety affecting the F atom when in proximity.  The C-site tr-XPS calculations in Fig. \ref{fig:CF}(d) confirm the lack of long-range Coulomb interaction due to all charges being localized around the methyl group. However, a 12 fs oscillation remains due to the umbrella motion. This oscillation appears to have double the frequency as the carbon is at the center of the oscillating system, contrary to the fluorine (see also animations of the dissociation dynamics in the SM). For delays below 25 fs, a fast chemical shift is observed due to the charge flowing from the hydrogen atoms to the carbon center [see black and green curves in Fig. \ref{fig:CF}(a)]. This is expected when the C-F bond breaks, as the hydrogens do not compete with the F atom so charge can flow from them to the C site.

From these measurements and calculations we demonstrate that even the relatively simple dissociation channel CH$_3$F$^+ \rightarrow$ CH$_3^+$+F illustrates the rich photochemical features that tr-XPS is able to discern. It also confirms the validity of extending the simple PC model (Eq. \ref{eq:partialchargemodel}) to the ultrafast regime. An interesting next step would be to confirm that this effect is also sensitive to coherent electron oscillations \cite{nisoli_attosecond_2017,rodriguez-cuenca_core-hole_2024}. It is worth noting that a similar effect related to distant Coulomb interactions was observed through ultrafast Auger spectroscopy \cite{wolf2017}. We now turn to further explore the sensitivity of the binding energies in the hydrogen dissociation channel [middle panel in Fig. \ref{fig:scheme} (a)].

\begin{figure*}\centering\includegraphics[scale=1.3]{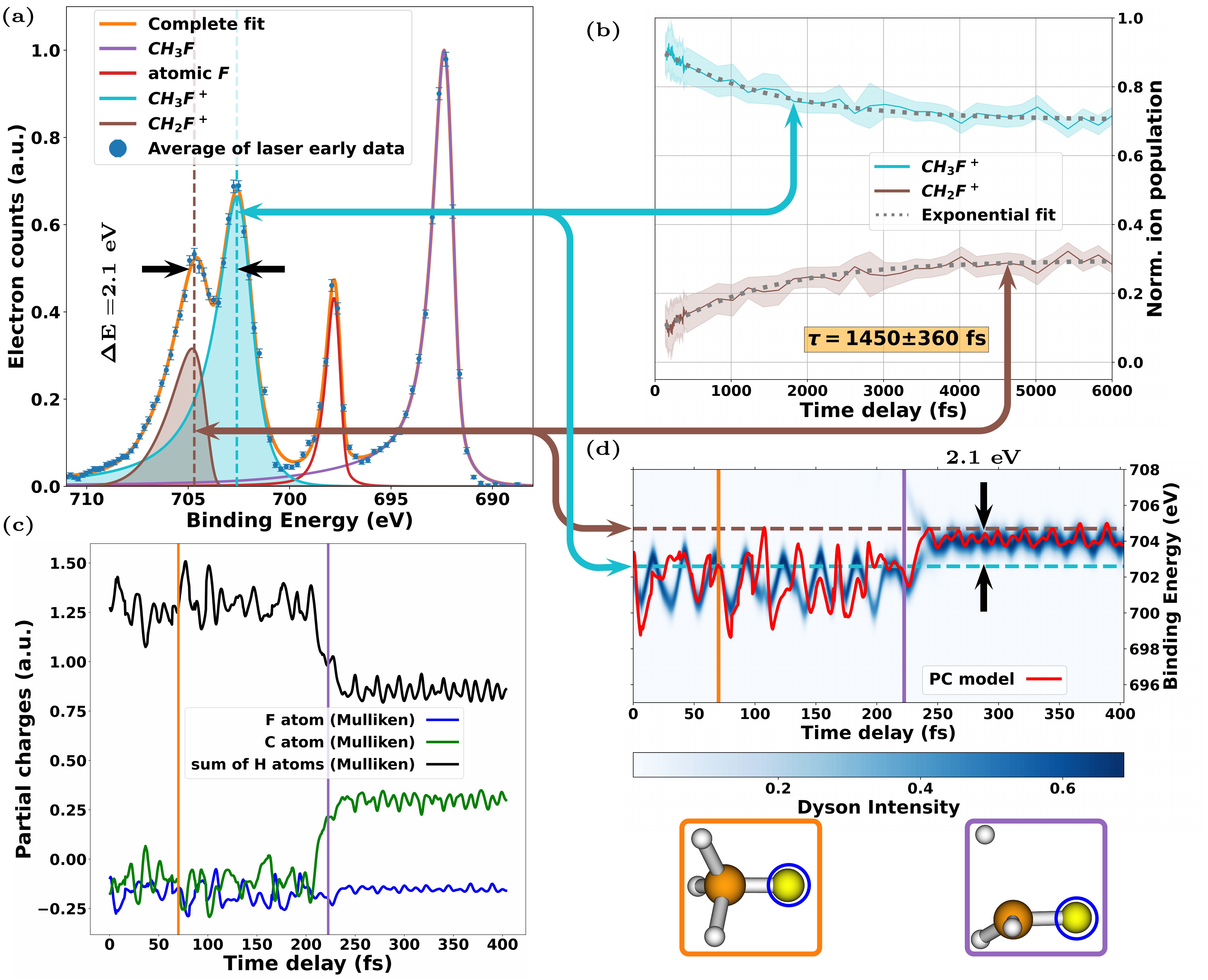}
\caption{{\bf Effects of the hydrogen detachment on the ultrafast chemical shifts.} a) Photoelectron spectra averaged between 2800 and 6000 fs and fitting of pseudo-Voigt profiles of the individual measured peaks (error bars represent the standard deviation). CH$_3$F$^+$ and CH$_2$F$^+$ are highlighted with shaded areas (cyan and brown respectively) and show a binding energy difference of 2.1 eV (see section S1-C in the SM for additional information on the data treatment and fitting). b) Time-evolution of the CH$_3$F$^+$ (cyan) and CH$_2$F$^+$ (brown) peaks showing the C-H bond dissociation rate for the CH$_3$F$^+$ cation, fitted with an exponential with a  decay constant of 1450 $\pm$ 360 fs (grey dotted line)  (c) Calculated Mulliken partial charges at the location of the F, C and H atoms, during a selected trajectory of C-H bond cleavage at 225 fs. For visualization purposes, we now use the sum of the charge of the H atoms instead of the average, as done in Fig. 2(a). (d) Calculated binding energy based on a PC model (red solid line), and dyson intensity calculated through the ab-initio approach including core-hole states (blue colorscale). Orange and purple lines in (c) and (d) mark the delays for the representative framed snapshots shown below (d). Similar to fig. \ref{fig:CF}(d), a weak satellite feature is observed at larger binding energies between 225 and 250 fs. This feature is also present at shorter time delays, but at binding energies beyond 708 eV (see figure S15 in the SM). }

\label{fig:CH}
\end{figure*}

\section*{tr-XPS sensitivity to hydrogen loss}

The hydrogen detachment from the parent molecule shows a noticeable chemical shift on both the C and F sites. This is corroborated by ab-initio calculations performed in the static molecules CH$_3$F$^+$ and CH$_2$F$^+$ that are in excellent agreement with the experimental data [see dashed blue lines in Figs. \ref{fig:scheme} (b) and (c)]. From the fluorine tr-XPS experimental trace we observe that the CH$_2$F$^+$ peak slowly increases over an extended period of time. To quantify the evolution of the populations we first perform a fitting procedure of pseudo-Voigt profiles \cite{schmid_new_2014} to the photoelectron spectra at later delays (average between 2800 and 6000 fs), where the population no longer evolves. This is shown in Fig. \ref{fig:CH}(a) (further details on the analysis procedure are provided in section S1-C in the SM). In summary, we first determine the contribution of each species to the overall shape of the measured F K-shell photoelectron spectrum. Then we proceed to fit the evolution in time for the individual CH$_3$F$^+$ and CH$_2$F$^+$ peaks, based on the measured tr-XPS trace. The result is shown in Fig. \ref{fig:CH}(b). We observe that the decrease of the CH$_3$F$^+$ and the increase of CH$_2$F$^+$ can be associated to a time constant of 1450 $\pm $ 360 fs and the relative populations converge to a ratio of 7:3 (in this we have neglected the minor channel of CHF$^+$ + H$_2$, see section S3-C in the SM for further details on this assumption).

To understand this dissociation process, we perform semi-classical ab-initio simulations of the dynamics. We first observe that the cation ground state may dissociate to CH$_2$F$^+$ and a hydrogen atom (see animations of some example trajectories showing these dynamics in the SM). From the potential energy curve calculated by stretching the C-H bond, we observe a small energy barrier, hence only if the cation molecule has enough energy, it may end up dissociating (see section S3-C in the SM). From the Franck-Condon geometry of the ground state, the required excess energy to reach dissociation is about 0.20 eV, and this dissociation occurs on slower time-scales than the one described in the previous section. 

We simulate one thousand semi-classical trajectories and we observe how some trajectories show the elongation and cleavage of the C-H bond at different points in time (see Fig. S13 in SM). This is similar to the behavior reported in investigations studying the roaming motion of hydrogens in molecules such as formaldehyde \cite{townsend_roaming_2004}. The partial charges calculation for one such trajectory is shown in Fig. \ref{fig:CH}(c) (hydrogen detachment occuring at 225 fs). To ensure all lines are clearly visualized, we use the sum of the charges on the hydrogen atoms rather than their average, as done in Fig. \ref{fig:CF}(a). We first notice that the partial charges at each site differ from the ones presented in Fig. \ref{fig:CF}(a), even at short time scales before the nuclear motion starts. This is due to the trajectories involving different excited states of the cation. We also observe that when the hydrogen is attached to the molecule, the positive charge is mainly localized on the methyl group, as F has a high electron affinity. The positive charge is then shared between the hydrogens and the carbon, but when one neutral hydrogen atom departs from the parent molecule, the local positive charge at the carbon site increases. Additionally, we observe a reduction in the amplitude of the oscillations on all atoms, due to the leaving hydrogen taking part of the nuclear kinetic energy.

As seen so far, after the loss of the hydrogen the increase of local charge around the C site leads to a chemical shift at both the C site and F site. Such chemical shifts after real-time charge redistribution from a local site have been reported before \cite{al-haddad_observation_2022,mayer_following_2022}, but interestingly a chemical shift at the F site occurs despite the partial charges being barely altered.  This behavior can also be attributed to the influence of surrounding atoms and their Coulomb interactions with the absorbing site, consistent with the findings presented in the previous section.  

As before, we compare the binding energies obtained using the PC model given by Eq. (\ref{eq:partialchargemodel}) with those derived from ab-initio calculations that explicitly include core-hole states to simulate the tr-XPS spectrum [see Fig. \ref{fig:CH}(d)]. While the model does not perfectly replicate the results of the core-hole calculations, it successfully captures the observed chemical shifts with notable accuracy. The discrepancies are likely due to the known limitations of assigning partial charges at hydrogen sites using Mulliken charge analysis \cite{saha_are_2009} (further discussion on the PC model and partial charge assignment is provided in Section S3-D of the SM).


\section*{Discussion}

We demonstrate that the combination of high spectral and temporal resolution in tr-XPS enables the real-time tracking of the dynamics of multiple photoproducts. Additionally, this technique can be extended to more complex systems and dynamics by employing a PC model to interpret the observed chemical shifts, particularly in cases where advanced binding energy calculations are computationally prohibitive. Notably, our results highlight the sensitivity to charges located around the x-ray absorbing site. This influence manifests in the experimental tr-XPS spectra as pronounced shifts following hydrogen detachment and, over longer distances, during C–F bond cleavage. We expect this effect to be a general phenomenon and to play a major role in the analysis of charge and proton transfer processes investigated with tr-XPS \cite{lennox_excited-state_2017,chen_amino_2018, douhal_proton-transfer_1996,schultz_efficient_2004,Neppl2015,oberli_site-selective-induced_2019,gu_chemical_2024}. Also, the measurement of core-level shifts resulting from a neutral atom or ion moving out of the Coulomb field of another ion opens up new possibilities for experiments on ionic dissociation.

The application of the PC model holds great promise for extending tr-XPS to investigate systems of higher complexity. In our study, Mulliken charge analysis proved sufficient to capture the main time-dependent spectral features, however, it highlighted the known limitations of the model. This could be improved through alternative approaches that are less sensitive to the choice of basis set, such as Natural Population Analysis (NPA) and Hirshfeld or Bader charge analysis \cite{cramer_essentials_2005}. Finally, an additional avenue for future investigation is the role of the Coulomb interaction between the core orbital and the valence orbitals in Eq. (\ref{eq:partialchargemodel}) in an evolving system.

\section*{Methods}

\subsection{Experimental Setup}

The optical pump laser \cite{palmer_pumpprobe_2019} is operated at a central wavelength of 800 nm and a pulse duration of 15 fs (FWHM). 1 mJ of energy is focused at the interaction region, in a spot of 80 $mu$m, leading to an intensity of approximately $5.5\times10^{14} \, W/cm^2$ in the interaction region. The pulses are delivered in bursts with a repetition rate of 10 Hz, each burst containing 81 pulses at an internal repetition rate of 188 kHz, leading to an effective repetition rate of 810 Hz. 

The x-ray FEL pulses are linearly horizontally polarized and centered at 928.3 eV photon energy, and monochromatized to a bandwidth of 0.33 eV (FWHM) via the SASE3 monochromator \cite{gerasimova_soft_2022}. The pulses are focused via a Kirkpatrick-Baez mirror pair \cite{mazza_beam_2023} to a few-$\mu$m spot in the interaction region, ensuring that the probing occurs within the region of highest intensity. The FEL is operated at the same repetition rate as the optical laser and both are optically synchronized \cite{schulz_femtosecond_2015}. 

The relative time delay between the pulses is controlled via a delay line in the optical laser arm, and is further corrected based on the measured arrival times of the individual electron bunches before the undulator modules. After correction, a temporal resolution of 35 fs is obtained (see Fig. S5 in the SM for further details), mainly limited by the x-ray pulse duration after the monochromator and possible long-term drift, unaccounted for by the electron arrival times \cite{rivas_high-temporal-resolution_2022}. 

X-ray photoelectrons are detected via a time-of-flight electron spectrometer detecting in the horizontal plane, parallel to the FEL polarization \cite{de_fanis_high-resolution_2022}. The laser is polarized perpendicular to the FEL, though some possible depolarization accumulated in the beam-delivery line might contribute to the observed laser-dressing effect.  

The CH$_3$F molecules are delivered to the interaction point in a fluoromethane/He gas mixture, with a 3\% concentration. Background pressure in the chamber reached up to $9\times 10^{-7}$ mbar when delivering the gas mixture. At these moderately high pressures and laser intensity there is measurable space charge, which is accounted for in the analysis (see supplementary material), but does not broaden the measured photolines.

\subsection{XPS static calculations at the dissociation limit}

In our model, the electronic structure of the cationic states $\mathrm{CH_3F}^+$ (with a hole in the valence shell), the core-hole states $\mathrm{C(1s^{-1})H_3F^{2+}}$ and $\mathrm{CH_3F(1s^{-1})^{2+}}$ as well as the satellite states (created when valence-to-valence excitation occurs simultaneously to core-ionization) is calculated at the complete active space self-consistent field method (CASSCF). The active space CASSCF(11,10) consists of 11 electrons distributed in 10 molecular orbitals (see section S2 of SM for more details). The ANO-L basis set is used together with a state average of 9 cationic states. Both the energy and gradient calculations at CASSCF level are performed with the OpenMolcas quantum chemistry package \cite{fdez_galvan_openmolcas_2019}. The XPS spectra shown in Fig. \ref{fig:scheme} are calculated at the dissociation limit with an ANO-L basis set, and the energies are corrected at the extended multistate complete active space at second order of perturbation theory (XMS-CASPT2). The XMS-CASPT2 calculations are in excellent agreement with the experimental spectral features.

\subsection{Ab-initio calculations for the photoinduced dynamics}

We develop a numerical approach based on a semi-classical surface-hopping model. In this model, the electronic structure of the molecule is treated at the quantum level, while the nuclei follow the quantum potential created by the electrons using a swarm of classical trajectories that are propagated using the Newton's equations of motion (see section S3-A in the SM for more details). At each time step of the trajectory, we perform the electronic calculations of the energies and the energy gradients on the fly. We use the semi-classical propagation implemented in SHARC \cite{richter_sharc_2011}. The classical nuclear trajectories are only subjected to a single electronic potential. In the regions of strong nonadiabatic couplings, the system may experience a change of electronic potential, and the probability to jump is a stochastic process that is evaluated at each nuclear time step from the change of the electronic populations \cite{petersen_electronic_2012}. 

We perform the electronic calculations at the CASSCF(11,10) level of theory and corrected using XMS-CASPT2. Most of the calculations (all the ones reported in the main text) were performed with the ANO-L basis. However, in order to reduce the computational cost we use the augmented correlation-consistent polarized valence double-zeta aug-cc-pVDZ basis set of Dunning (instead of the ANO-L basis) for some of the calculations reported in section S3-D of the SM. This basis set shows a good agreement with the ANO-L basis set (see section S3-C of the SM).

The initial coordinates and velocities for the trajectories are obtained following an harmonic Wigner distribution in the ground electronic state of the neutral molecule. The strong-field ionization of the IR pulse is modeled by projecting the ground state distribution into the three lowest cationic states, where they are then propagated.

The hydrogen detachment process depends on the initial distribution and we therefore need a large amount of trajectories to obtain statistics and compare it with the experimental data. The ground state of the cation has a small energy barrier to reach the H detachment and only trajectories with enough vibrational energy may end up losing a hydrogen. The energy barrier is about 0.20 eV from the Frank-Condon region, and 0.84 eV from the minimum energy of the cation ground state. We simulate the dynamics in the ground state of the cation with SHARC, calculating the electronic structure at the Coupled-Cluster Single and Double (CCSD) level using the Psi4 program \cite{turney_psi4_2012}. The CCSD calculations enables us to speed up the simulations and perform a large amount of trajectories, and the energies are in good agreement with XMS-CASPT2 calculations and previous experimental data \cite{locht_mass_1987} (see section S3-C of the SM for more details).

The partial charges employed in the partial charges model of Eq. (\ref{eq:partialchargemodel}) were computed employing the data of a Mulliken partial charge analysis obtained from the simulations of the dynamics. We corrected the charges through a fit of the transient XPS signals at carbon and fluorine K-shell (see section S3-D of the SM for more details) in order to avoid an overestimation of the hydrogen partial charges \cite{saha_are_2009}.

\subsection{Ab-initio calculations for time-resolved XPS}

The core-hole states are calculated at the CASSCF(12,11) level, where the 1s orbital from the ionized electron is included in the active space. We compute a state average over 10 states in order to include the contribution of the main core hole and other relevant satellite states. The binding energies are calculated by taking the difference of the initial states and the corresponding core-hole states. To compute the time-resolved XPS spectra, we evaluate the ionization amplitudes at each step of the propagation from the Dyson orbitals \cite{ortiz_dyson-orbital_2020}. The total time-resolved XPS spectra are calculated by doing an incoherent sum over the ionization amplitudes. In order to take into account the x-ray probe pulse bandwidth, we apply a convolution of the spectra with a Gaussian of 0.26 eV (FWHM) bandwidth.

\section*{Acknowledgments} The authors acknowledge the support groups of
European XFEL and DESY for their assistance during these experiments. We additionally acknowledge comments by Serguei Molodtsov during the preparation of this manuscript. L.P. and A.P. acknowledge the Spanish Ministry of Science, Innovation and Universities \& the State Research Agency through grants refs. PID2021-126560NB-I00 and CNS2022-135803 (MCIU/AEI/FEDER, UE), and the "Mar\'ia de Maeztu" Programme for Units of Excellence in R\&D (CEX2023-001316-M), and computer resources and assistance provided by Centro de Computaci\'on Cient\'ifica de la Universidad Aut\'onoma de Madrid and RES resources (FI-2024-3-0011, FI-2024-2-0034, FI-2023-2-0012, FI-2022-3-0022, FI-2022-1-0031). S.O ackowledges support from the Swiss National Science Foundation, National Center of Competence in Research - Molecular Ultrafast Science and Technology NCCR - MUST. T.M. and M.M. acknowledge support from the DFG, the German Research Foundation (Deutsche Forschungsgemeinschaft, Project 170620586, SFB-925). MI acknowledges support by the Bundesministerium für Bildung und Forschung (BMBF) Grant No. 13K22CHA, the Cluster of Excellence “Advanced Imaging of Matter” of the Deutsche Forschungsgemeinschaft (DFG)—EXC 2056—Project ID No. 390715994, and (DFG)—Project No. 328961117-SFB 1319 ELCH (“Extreme light for sensing and driving molecular chirality”). DRo was supported by the National Science Foundation grant no. PHYS-2409365. AR was supported by the Chemical Sciences, Geosciences, and Biosciences Division, Office of Basic Energy Sciences, Office of Science, US Department of Energy, grant no. DE-FG02-86ER13491. This publication is based upon work from COST Action NEXT, CA22148 supported by COST (European Cooperation in Science and Technology).

\section*{Data availability}
Data underlying the results presented is available under request at https://doi.org/10.22003/XFEL.EU-DATA-002730-00.

\clearpage

\bibliography{referencesPRX}
\bibliographystyle{unsrt}


\end{document}